# Quantum Electrodynamics of Nanosystems

Samina S. Masood

Physics Department, University of Houston Clear Lake

2700 Bay Area Blvd, Houston, TX 77058, USA; masood@uhcl.edu

## ABSTRACT

Quantum description of mulitiparticle nano-systems is studied in a hot and dense electromagnetic medium. We use renormalization techniques of quantum field theory to show that the electromagnetic properties like electric permittivity and magnetic permeability depend on the temperature and density of the media. Casimir force also depends upon the physical properties of the medium and becomes a function of these parameters within the nano-systems. We discuss the effect of the Casimir force on the nanosystems in terms of temperature and density of the system. We present carbon nanotubes and biomolecules as examples.

**Keywords:** *Nanosystems, Casimir Force, Protein Molecules, Nanotubes,*

## 1. INTRODUCTION

The electromagnetic description of chemical processes and its application to biochemistry is used for a long time to understand the complicated structures and growth of living cells. Water being the most abundant solvent plays a big role in the growth and survival of life. A majority of the chemical reactions involve polar molecules dissolved in water. In this process, they release $H^+$ or $OH^-$ that can affect the acidic properties of the medium. The change in pH value affects the cell growth [1-4] directly. However, the lipid walls may develop Casimir [5] type force, if it becomes within an appropriate distance with the non-polar molecules. Not only that, the entanglement in the carbon nanotubes the hydrophobic part of the protein molecules may also develop accidental Casimir force that can lead to unexpected behavior.

Casimir force is known to depend on the electromagnetic properties of the medium. It is simply proportional to the electric permittivity of the medium [6]. It has also been calculated that temperature [7] affects the Casimir force. Protein densities [8] are believed to be between (1.22-1.47) g/cc. The local densities at certain ends of protein molecule may be large enough to include the density effects. Therefore the vacuum polarization and other radiative corrections are included in this study. We show the dependence of Casimir force on temperature and density using the renormalization scheme of quantum electrodynamics (QED). The brief details of the calculations are given in the next section. We devote section 3 to discuss QED of Nanosystems. In the last section we will present some of the applications of these results to Nanosystems.

## 2. MEDIUM EFFECTS

We consider Nanosystems as multi-particle QED systems in a theromodynamic equilibrium background. It is believed that the QED effects are non-ignorable on nanoscales and the perturbative effects of QED can add up to sizeable corrections. Renormalization techniques are a way to extract finite perturbative contribution to physically measureable quantities in quantum field theory.

Second quantization in QED leads to the calculation of vacuum polarization at nanoscales to prove renormalization of QED as a gauge theory and remove the unwanted singularities to prove it as a physical theory. Electromagnetic properties of the medium including the electric permittivity $\epsilon_E$ and the magnetic permeability $\mu_B$ are calculated from the vacuum polarization of photon. We have previously shown [9-15] that the vacuum polarization at the one loop level does not affect the QED coupling at low temperatures. However we calculate nonzero effects to $\epsilon_E$ and $\mu_B$ at extremely low temperatures if the chemical potential $\mu$ is larger than temperature.

We showed [9] that the electromagnetic properties are modified as a function of temperature and chemical potential of the media regardless of their values at the higher loop level though the contribution may be too small to be worried about. It basically happens that the longitudinal and the transverse components of the polarization tensor are not the same in this situation.

The one loop radiative corrections to the vacuum polarization tensor are proportional to the coupling constant $\alpha_R$ which can be renormalized in hot and dense media to pick up the effect of temperature and density on the QED coupling constant at low temperature and small values of density are calculated [9] as,

$$\alpha_R = \alpha(T=0)\left(1 + \frac{\alpha^2 T^2}{6m^2}\right).$$

At high temperatures (more than millions of Kelvin) the one-loop corrections are non-negligible but those temperatures are relevant to the very early universe or the stellar cores and are out of the scope of this paper. However, the $\epsilon_E$ and the $\mu_B$ of a highly dense medium are modified with temperature and chemical potential. It has pronounced effects at large value of density or chemical potential $\mu$. Much larger values of $\mu$ correspond to local densities of large organic molecules like proteins that may be much larger than the average protein densities [8]. The mass of electron itself is modified in the highly dense background of organic molecules at nanoscales that is almost a local effect in that region. The change in the electron mass m has been calculated [10] as a function of chemical potential as

$$\frac{\delta m}{m} \simeq -\frac{3\alpha}{\pi} \ln \frac{\mu}{m} + \frac{\alpha}{\pi} \left[1 - \frac{\mu^2}{m^2}\right] \left[3\frac{m^2}{\mu^2} + 2\frac{p^2}{\mu^2} - 1\right].$$

This extremely large local density will have an associated local magnetic field that will couple with the charged particles of the system including electrons, polar atoms and the molecule through the magnetic moment [11,12] depending on the strength of the magnetic field.

$$\mu_a^{el} \simeq \frac{\alpha}{2\pi} - \frac{2\alpha\pi}{9} \frac{T^2}{m^2} - \frac{4\alpha}{3\pi m^2} \left[ I'_1 - \frac{3m^2}{2} I'_2 - \frac{m^4}{8} I'_3 \right]$$

Whereas $I'$ are complicated functions of chemical potentials. The details of these functions can be found in Ref. [12]. A very simple form of these integrals at $T \ll \mu$ is given by

$$I'_1 = \frac{\mu^2}{2} \left[1 - \frac{m^2}{\mu^2}\right],$$

$$I'_2 = \ln(\mu/m),$$

$$I'_3 = \frac{1}{2m^2}(1 - m^2/\mu^2).$$

The magnetic moment of other charged particles like atoms or molecules may not get sizeable corrections from background and their intrinsic magnetic dipole moment may be ignorable also.

The high local density inside the large and complicated organic molecules will affect the vacuum polarization in the non-polar portion of molecules. The local values of electric permittivity $\epsilon_E$ and the magnetic permeability $\mu_B$ of certain regions may be expressed [10] in terms of the plasma frequencies of photons $\omega$ and the wave-vector k. The details of these calculations can also be found in Ref. [10]. Local values of these quantities are the ones which may fluctuate between very high and very low values significantly due to the free energy of the system.

The $\epsilon_E$ and $\mu_B$ are evaluated as functions of chemical potential of the system as,

$$\epsilon_E(K,\mu,T) \simeq 1 - \frac{2\alpha}{\pi^2 K^2 k} \left[ \left[k - 2w \ln \frac{w-k}{w+k}\right] \mathcal{I}_1 \right.$$
$$\left. - 2A_1 \mathcal{I}_2 + \frac{1}{2} A_2 \mathcal{I}_3 \right] \left[1 - \frac{w^2}{k^2}\right]$$

And

$$\frac{1}{\mu_B(K,\mu,T)} \simeq 1 + \frac{2\alpha}{\pi k^3 K^2} \left[ \left[\frac{k^2}{2} + w^2\right] \left[1 - \frac{w^2}{k^2}\right] \left\{2w \ln \frac{w-k}{w+k} \mathcal{I}_1 - 2A_1 \mathcal{I}_2 - \frac{1}{2} A_2 \mathcal{I}_3\right\} \right.$$
$$\left. - \frac{k^3 K^2 + k^4 m}{4} \mathcal{I}_2 - \frac{m^2}{2} \left\{\frac{k^5}{12} + \frac{(3w^2+k^2)k^3 K^2}{24m^2} + \frac{m^2 k^2}{8k}(5w^2 - 3k^2)\right\} \mathcal{I}_3 \right]$$

This difference arises because of the fact that the vacuum polarization tensor indicates different polarization along the line of separation of conductors and the perpendicular components. $\pi_L$ is along the direction of separation of two conductors and $\pi_T$ is perpendicular to the separation of conductors. The longitudinal and transverse components of the vacuum polarization tensor are affected differently by the temperature and densities of the medium and are expressed as a function of T and $\mu$ of the system are given as

$$\Pi_L(0,\mathbf{k}) \equiv K_L^2 \simeq \frac{e^2}{2\pi^2} \left[\mathcal{I}_1 + 2m^2 \mathcal{I}_2 - \frac{9m^4}{2} \mathcal{I}_3\right],$$

$$\Pi_T(0,\mathbf{k}) \equiv K_T^2 \simeq \frac{e^2}{2\pi^2}(m^2 \mathcal{I}_2 - \tfrac{17}{4} m^4 \mathcal{I}_3),$$

Whereas,

$$\mathcal{I}_1(T,\mu) = \tfrac{1}{2} \int E \, dE [n_F(E+\mu) + n_F(E-\mu)]$$
$$= \frac{1}{2} \left[ \frac{\mu^2}{2}\left[1 - \frac{m^2}{\mu^2}\right] + \frac{1}{\beta}\{a(m\beta,\mu) - a'(m\beta,\mu)\} \right.$$
$$\left. - \frac{1}{\beta^2}\{c(m\beta,\mu) + c'(m\beta,\mu)\} \right],$$

$$\mathcal{I}_2(T,\mu) = \tfrac{1}{2} \int \frac{dE}{E} [n_F(E+\mu) + n_F(E-\mu)]$$
$$= \frac{1}{2} \left[ \ln \frac{\mu}{m} + b(m\beta,\mu) + b'(m\beta,\mu) \right],$$

$$\mathcal{I}_3(T,\mu) = \frac{1}{2} \int \frac{dE}{E^3} [n_F(E+\mu) + n_F(E-\mu)]$$
$$= \frac{1}{2} \left[ \frac{1}{2m^2}\left[1 - \frac{m^2}{\mu^2}\right] - \frac{1}{4\mu^2} + \frac{1}{m^2}\{n_f(\mu+m) + n_F(\mu-m)\} \right.$$
$$\left. + \frac{\beta}{m} \left\{\frac{e^{-\beta(\mu+m)}}{[1+e^{\beta(\mu+m)}]^2} + \frac{e^{-\beta(\mu-m)}}{[1+e^{-\beta(\mu-m)}]^2}\right\} + d(m\beta,\mu) + d'(m\beta,\mu) \right.$$

At the expected local densities of organic molecules they can simply be written as (for i = 1,2, or 3)

$$\mathcal{J}_i = \tfrac{1}{2} I'_i$$

Using $\pi_L$ and $\pi_T$, the real and imaginary parts of the dielectric function and the absorption coefficient of carbon Nanotubes can also be shown to be a function of temperature and density. In the real physical Nanotubes, the structural defects and the un-expected entanglement may be due to these probabilistic effects. Casimir effect may be a big cause of this entanglement which is as random and difficult to explain as the Casimir force is.

This difference between the longitudinal and the transverse polarization is a possible source of Casimir type force between two conducting plates. This force may be attractive or repulsive depending on the possibility of facing like or unlike charges. This kind of force can exist between neutral ends of molecules may locally act as small conductors.

## 3. OTHER QED EFFECTS

The basic electromagnetic processes that control the dynamics of polar and non-polar molecules in water based fluid in living cells and determine the growth and the survival probability of life. Two basic interactions of solute molecules in water solvent are *hydrophilic* (water-loving) and *hydrophobic* (water-fearing) interactions. The study of these interactions with this approach may resolve some basic issues of chemical synthesis and even may lead to methods to develop more efficient energy sources.

Generally, hydrophilic molecules carry polar and/or ionized functional groups whereas hydrophobic molecules are electrically neutral with symmetric carbon chains. When non-polar molecules like lipids are mixed with water, they move away from water molecules and end-up gathering at the surface of water through hydrophobic interaction. Another good example of hydrophobic interaction is the addition of oil to water that gives two distinct layers of non-polar and polar liquid. Large molecules with a polar/ionized region and one end and a non-polar region at the other end have both hydrophilic and hydrophobic characteristics. The increasing solubility of these molecules in water is believed to be due to the cluster formation on the hydrophilic region and the hydrophobic pockets on the other side. The physics of these interactions is not so well-understood. We are using quantum electrodynamics (QED) to properly describe these hydrophobic and hydrophilic interactions. Even the oral consumption and/or inhaling of chemical components like alcohol or other charged species will change the electrodynamics of living bodies. It may even effect the magnetic moments of molecules.

The effect of electromagnetic radiation as a source of external energy can also be expected to have tremendous effects on human body. The quantum mechanical uncertainties and the probability theory may explain the causes of gene mutation [16] that can better describe the cause of developing cancer in human bodies. It is so much dependent on the dosage and location that the radiation therapy is used to control the cancer cell growth as the radiation provides energy which shields the cell nutrition and may reduce the growth. The affect on the pH value of the nutrient also plays a role in the cell growth.

Another type of QED effect is possible due to the inhomogeneous magnetic fields which may induce Eddy currents in different places of electrolytes that serve as nutrient and effect the cell growth. These currents are extremely small and ignorable individually, however for large molecules they may add up to sizeable effect. Especially because they may cause the molecular movements in preferred directions and effect the cell mechanism, its consumption of charged radicals like calcium ions or potassium ions and ultimately effect the cell growth. Analytical calculations of these currents are too cumbersome even the existing computational models cannot handle it. So the best way to study the importance of these effects is the experimental study which can lead to the theoretical explanation of those results. These Eddy currents can change the membrane potential and even the channel potentials. External weak magnetic fields will couple with the magnetic moment of molecules at different level depending on the strength and may give measurable effects. We have been working on some very simple experiments to just check the plausibility of this approach.

## 4. **DISCUSSION OF RESULTS**

The temperature contributions to the QED parameters are not important for Nanosystems at all so we do not need to study them in detail. However, we need to use the thermodynamic principals to quantify the chemical potential effects in Nanosystems.

Casimir force may be the accidently generated force that may lead to the unexpected and unexplained behavior of the protein folding and carbon Nanotubes. Not only that, dense localities of these molecules like hydrophobic portion of protein molecules may have high densities and chemical potentials instantaneously to generate stronger than expected Casimir force.

The dependence of electric permittivity and magnetic permeability on temperature and chemical potential changes the electromagnetics of thermodynamic media. This is a possible reason of change of dielectric properties of the medium as well as the displacement vectors in the long conducting chains of organic molecules or some portions of carbon Nanotubes or protein molecules.

In carbon nanotubes, instantaneous Casimir force gives a possible explanation of synthetic defaults like entanglement of carbon Nanotubes, etc.

We are mainly proposing to use basic concepts of perturbative QED such as dipole moment of molecules, their interaction with electromagnetic fields, eddy currents produced through the variable magnetic fields and especially the possible Casimir forces on the non-polar end of the large molecule and the Lamb shift.

## REFERENCES :

1. See for example: S. Humez et al.`The role of intracellular pH in cell growth arrest induced by ATP' *Am J Physiol Cell Physiol* 287: C1733-C1746, (2004).
2. M. Mataragas, `Influence of nutrients on growth and bacteriocin production by *Leuconostoc mesenteroides* L124 and *Lactobacillus curvatus* L442' , A ntonie van Leeuwenhoek **85:** 191–198, (2004).
3. H. `Umashankar, et. al., Influence of nutrients on cell growth and xanthan production by Xanthomonas campestris' Bioprocess Engineering 14 307-309 (1996).
4. Dan Zilberstein, et al;, Escherichia coli Intracellular pH, Membrane Potential, and Cell Growth, J. of Bacteria, , 246-252 (1984); Sarah Sundelacruz, Michael Levin and David L. Kaplan, `Role of Membrane Potential in the Regulation of Cell Proliferation and Differentiation, Stem Cell Rev and Rep 5: 231–246 (2009).
5. Casimir, H. G. B. "On the attraction between two perfectly conducting plates." *Proc. Con. Ned. Akad. van Wetensch* B51 (7): 793-796 (1948). Lamoreaux, S. K. "Demonstration of the Casimir force in the 0.6 to 6 mm range." *Physical review Letters* 78 (1): 5-8 (1997).
6. K. A. Milton, Invited Lectures at 17th Symposium on Theoretical Physics, Seoul National University, Korea, June 29 - July 1, (1998); http://arxiv.org/PS_cache/hep-th/pdf/9901/9901011v1.pdf; and references therein.
7. Simen Å. Ellingsen, Iver Brevik, Kimball A. Milton,` Casimir effect at nonzero temperature for wedges and cylinders', Phys.Rev.D81: 065031, (2010). And references therein.
8. Pieter Rein ten Wolde and Daan Frenkel,` Enhancement of Protein Crystal Nucleation by Critical Density Fluctuations' SCIENCE VOL. 277 Spt. 1797O9.; H. Fischer, Igor Polikarpov and Aldo F.Craievich, `Average protein density is a molecular-weight dependent function', Protein Sci. 2004 October; 13(10): 2825–2828 and several other papers.
9. **Samina Masood** and Mahnaz Haseeb,`Second Order Corrections to QED Coupling at Low Temperatures` *Int. J. Mod. Phys*., A23 (4709-4719), 2008. ; **Samina Masood** and Mahnaz Haseeb,`Second Order Thermal Corrections to Electron Wavefunction' Phys.Lett.B704:66-73,2011.
10. **Samina Masood**, Renormalization of QED in Superdense Media', Phys. Rev. D47. (648-652) 1993. And references therein.
11. K.Ahmed and **Samina Masood**,`Vacuum Polarization Effects at Finite Temperature and Density in QED', *Ann. of Phys. (*N.Y), 207 , (460-473), 1991;. **Samina Masood**, `Finite Temperature and Density Effects on Electron Self-mass and Primordial Nucleosynthesis', Phys. Rev. D36, 2602 (1987).; K.Ahmed and **Samina Masood**, `Finite Temperature and Density Renormalization Effects in QED', Phys. Rev.D35, 4020 (1987); K.Ahmed and **Samina Masood**, `Renormalization and Radiative Corrections at Finite Temperature Reexamined', Phys. Rev.D35, 1861 (1987),
12. **Samina Masood**, `Photon mass in the classical limit of finite temperature and density', Phys. Rev. D 44, 3943–3948 (1991).
13. **Samina Masood**, `Study of Casimir Effect in Nanosystems', BAPS.2010. MAR.S1.254
14. **Samina Masood**, ` Casimir Effect in Carbon Nanotubes' A talk presented at the First International Symposium on Nanotechnology, energy and space, Oct. 2009 Houston.
15. Phu Nuygen, Mike Cabrera, Channing Moeller and **Samina Masood,** `Casimir Effect and its Applications to Biophysics' ,BAPS.2009. TSF.D1. 8. (Fall APS meeting of Texas section, Oct. 2009)
16. Bertram J "The molecular biology of cancer". *Mol. Aspects Med.* **21** (6): 167–223(2000).